\documentclass[twocolumn,apj]{emulateapj}
\usepackage{apjfonts}
\epsscale{1.15}

\shorttitle{A Strongly Lensed SMG at $z\sim3$} 
\shortauthors{Fu et al.}
\journalinfo{ApJ accepted}
\submitted{ApJ accepted}

\newcommand{\wise}{{\it WISE}}
\newcommand{\planck}{{\it Planck}}
\newcommand{\herschel}{{\it Herschel}}
\newcommand{\um}{$\mu$m}
\newcommand{\uJy}{$\mu$Jy}

\newcommand{\kms}{{km s$^{-1}$}}
\newcommand{\msun}{${\rm M}_{\odot}$}
\newcommand{\lsun}{${\rm L}_{\odot}$}
\newcommand{\msunyr}{$M_{\odot}~{\rm yr}^{-1}$}
\newcommand{\lir}{$L_{\rm 8-1000}$}

\def\geqsim{\lower.73ex\hbox{$\sim$}\llap{\raise.4ex\hbox{$>$}}$\,$}
\def\leqsim{\lower.73ex\hbox{$\sim$}\llap{\raise.4ex\hbox{$<$}}$\,$}

\newcommand{\objfull}{HATLAS~J114637.9$-$001132}
\newcommand{\obj}{HATLAS12$-$00}
\newcommand{\dra}{$\delta$RA}
\newcommand{\ddec}{$\delta$Dec}
\newcommand{\sersic}{S\'{e}rsic}
\newcommand{\K}{$K$}
\newcommand{\muKeck}{$16.7\pm0.8$}
\newcommand{\muSMA}{$7.6\pm1.5$}
\newcommand{\muJVLA}{$6.9\pm1.6$}
\newcommand{\mk}{}

\begin{document}

\title{A Comprehensive View of a Strongly Lensed Planck-Associated Submillimeter Galaxy}

\author{
Hai~Fu\altaffilmark{1}, E.~Jullo\altaffilmark{2}, A.~Cooray\altaffilmark{1}, R.S.~Bussmann\altaffilmark{3}, R.J.~Ivison\altaffilmark{4}, I.~P{\'e}rez-Fournon\altaffilmark{5,6}, S.G.~Djorgovski\altaffilmark{7,8}, N.~Scoville\altaffilmark{7}, L.~Yan\altaffilmark{7}, D.A.~Riechers\altaffilmark{7}, J.~Aguirre\altaffilmark{9}, R.~Auld\altaffilmark{10}, M.~Baes\altaffilmark{11}, A.J.~Baker\altaffilmark{12}, M.~Bradford\altaffilmark{7}, A.~Cava\altaffilmark{13}, D.L.~Clements\altaffilmark{14}, H.~Dannerbauer\altaffilmark{15,16}, A.~Dariush\altaffilmark{17}, G.~De~Zotti\altaffilmark{18,19}, H.~Dole\altaffilmark{20}, L.~Dunne\altaffilmark{21}, S.~Dye\altaffilmark{21}, S.~Eales\altaffilmark{10}, D.~Frayer\altaffilmark{22}, R.~Gavazzi\altaffilmark{23}, M.~Gurwell\altaffilmark{3}, A.I.~Harris\altaffilmark{24}, D.~Herranz\altaffilmark{25}, R.~Hopwood\altaffilmark{17}, C.~Hoyos\altaffilmark{21}, E.~Ibar\altaffilmark{4}, M.J.~Jarvis\altaffilmark{26,27}, S.~Kim\altaffilmark{1}, L.~Leeuw\altaffilmark{28,29}, R.~Lupu\altaffilmark{9}, S.~Maddox\altaffilmark{21}, P.~Mart{\'i}nez-Navajas\altaffilmark{6,5}, M.J.~Micha{\l}owski\altaffilmark{30}, M.~Negrello\altaffilmark{18,31}, A.~Omont\altaffilmark{23}, M.~Rosenman\altaffilmark{9}, D.~Scott\altaffilmark{32}, S.~Serjeant\altaffilmark{31} I.~Smail\altaffilmark{33}, A.M.~Swinbank\altaffilmark{33}, E.~Valiante\altaffilmark{10}, A.~Verma\altaffilmark{34}, J.~Vieira\altaffilmark{7}, J.L.~Wardlow\altaffilmark{1}, P.~van~der~Werf\altaffilmark{35}
}
\altaffiltext{1}{Department of Physics \& Astronomy, University of California, Irvine, CA 92697; haif@uci.edu}
\altaffiltext{2}{Observatoire d'Astrophysique de Marseille-Provence, 38 rue Fr\'ed\'eric Joliot-Curie, F-13388 Marseille, France}
\altaffiltext{3}{Harvard-Smithsonian Center for Astrophysics, 60 Garden Street, Cambridge, MA 02138}
\altaffiltext{4}{UK Astronomy Technology Centre, Royal Observatory, Edinburgh, EH9 3HJ, UK}
\altaffiltext{5}{Instituto de Astrof{\'\i}sica de Canarias (IAC), E-38200 La Laguna, Tenerife, Spain}
\altaffiltext{6}{Departamento de Astrof{\'\i}sica, Universidad de La Laguna (ULL), E-38205 La Laguna, Tenerife, Spain}
\altaffiltext{7}{California Institute of Technology, 1200 E. California Blvd., Pasadena, CA 91125}
\altaffiltext{8}{Distinguished Visiting Professor, King Abdulaziz Univ., Jeddah, Saudi Arabia}
\altaffiltext{9}{Department of Physics and Astronomy, University of Pennsylvania, Philadelphia, PA 19104}
\altaffiltext{10}{School of Physics and Astronomy, Cardiff University, The Parade, Cardiff, CF24 3AA, UK}
\altaffiltext{11}{Sterrenkundig Observatorium, Universiteit Gent, Krijgslaan 281 S9, B-9000 Gent, Belgium}
\altaffiltext{12}{Department of Physics and Astronomy, Rutgers, The State University of New Jersey, 136 Frelinghuysen Rd, Piscataway, NJ 08854}
\altaffiltext{13}{Departamento de Astrof\'{\i}sica, Facultad de CC. F\'{\i}sicas, Universidad Complutense de Madrid, E-28040 Madrid, Spain}
\altaffiltext{14}{Astrophysics Group, Imperial College London, Blackett Laboratory, Prince Consort Road, London SW7 2AZ, UK}
\altaffiltext{15}{Universit\"at Wien, Institut f\"ur Astronomie, T\"urkenschanzstra{\ss}e 17, 1160 Wien, \"Osterreich}
\altaffiltext{16}{Laboratoire AIM-Paris-Saclay, CEA/DSM-CNRS-Universit\'e Paris Diderot, Irfu/SAp, CEA-Saclay, 91191 Gif-sur-Yvette Cedex, France}
\altaffiltext{17}{Physics Department, Imperial College London, Prince Consort Road, London SW7 2AZ, UK}
\altaffiltext{18}{INAF - Osservatorio Astronomico di Padova, Vicolo dell'Osservatorio 5, I-35122 Padova, Italy}
\altaffiltext{19}{SISSA, Via Bonomea 265, I-34136, Trieste, Italy}
\altaffiltext{20}{Institut d'Astrophysique Spatiale (IAS), b\^atiment 121, Universit\'e Paris-Sud 11 and CNRS (UMR 8617), 91405 Orsay, France}
\altaffiltext{21}{School of Physics and Astronomy, University of Nottingham, NG7 2RD, UK}
\altaffiltext{22}{NRAO, PO Box 2, Green Bank, WV 24944}
\altaffiltext{23}{Institut d'Astrophysique de Paris, UMR 7095, CNRS, UPMC Univ. Paris 06, 98bis boulevard Arago, F-75014 Paris, France}
\altaffiltext{24}{Department of Astronomy, University of Maryland, College Park, MD 20742-2421}
\altaffiltext{25}{Instituto de F{\'i}sica de Cantabria (CSIC-UC), Avda. los Castros s/n, 39005 Santander, Spain}
\altaffiltext{26}{Centre for Astrophysics Research, Science \& Technology Research Institute, University of Hertfordshire, Hatfield, Herts, AL10 9AB, UK}
\altaffiltext{27}{Physics Department, University of the Western Cape, Cape Town, 7535, South Africa}
\altaffiltext{28}{Physics Department, University of Johannesburg, P.O. Box 524, Auckland Park, 2006, South Africa}
\altaffiltext{29}{SETI Institute, 189 Bernardo Ave, Mountain View, CA, 94043}
\altaffiltext{30}{Institute for Astronomy, University of Edinburgh, Royal Observatory, Blackford Hill, Edinburgh EH9 3HJ, UK}
\altaffiltext{31}{Department of Physical Sciences, The Open University, Walton Hall, MK7 6AA Milton Keynes, UK}
\altaffiltext{32}{Department of Physics \& Astronomy, 325-6224 Agricultural Road, University of British Columbia, Vancouver, BC V6T 1Z1, Canada}
\altaffiltext{33}{Institute for Computational Cosmology, Durham University, Durham DH1 3LE, UK}
\altaffiltext{34}{University of Oxford, Oxford Astrophysics, Denys Wilkinson Building, Keble Road, Oxford, OX1 3RH, UK}
\altaffiltext{35}{Leiden Observatory, Leiden University, P.O. Box 9513, NL-2300 RA Leiden, The Netherlands}

\begin{abstract} 

We present high-resolution maps of stars, dust, and molecular gas in a strongly lensed submillimeter galaxy (SMG) at $z = 3.259$. \objfull\ is selected from the \herschel-Astrophysical Terahertz Large Area Survey (H-ATLAS) as a strong lens candidate mainly based on its unusually high 500~\um\ flux density ($\sim$300~mJy). It is the only high-redshift \planck\ detection in the 130~deg$^2$ H-ATLAS Phase-I area. Keck Adaptive Optics images reveal a quadruply imaged galaxy in the \K-band while the Submillimeter Array and the Jansky Very Large Array show doubly imaged 880~\um\ and CO(1$\to$0) sources, indicating differentiated distributions of the various components in the galaxy. \mk{In the source plane}, the stars reside in three major kpc-scale clumps extended over $\sim$1.6~kpc, the dust in a compact ($\sim$1~kpc) region $\sim$3~kpc north of the stars, and the cold molecular gas in an extended ($\sim$7~kpc) disk $\sim$5~kpc northeast of the stars. The emission from the stars, dust, and gas are magnified by $\sim$17, $\sim$8, and $\sim$7 times, respectively, by four lensing galaxies at $z \sim 1$. Intrinsically, the \mk{lensed} galaxy is a warm ($T_{\rm dust} \sim 40-65$~K), hyper-luminous ($L_{\rm IR} \sim 1.7\times10^{13}$~\lsun; SFR $\sim 2000$~\msun~yr$^{-1}$), gas-rich ($M_{\rm gas}/M_{\rm baryon} \sim 70\%$), young ($M_{\rm stellar}/{\rm SFR} \sim 20$~Myr), and short-lived ($M_{\rm gas}/{\rm SFR} \sim 40$~Myr) starburst. With physical properties similar to unlensed $z > 2$ SMGs, \objfull\ offers a detailed view of a typical SMG through a powerful cosmic microscope.

\end{abstract}


\section{Introduction}

Bright submillimeter-selected galaxies \citep[SMGs][]{Blain02} provide a powerful probe into the distant Universe. Thanks to the negative $K$-correction in the Rayleigh-Jeans tail of the dust thermal emission, flux limited submillimeter surveys \mk{with 850~\um\ flux density $S_{\rm 850} > 5$~mJy} reach an almost uniform integrated infrared (IR) luminosity limit across a wide redshift range ($1 < z < 8$) and yield a galaxy population mostly at redshifts between $1.7 < z < 2.8$ \citep{Chapman05}. With star formation rates (SFRs) of $\sim10^3$ $M_{\odot}$~yr$^{-1}$, the SMGs are the most intense star-forming galaxies, despite their inevitably short-lived nature (lifetime $\lesssim$ 0.1~Gyr). Although such intense starburst systems are extremely rare in the local Universe, SMGs and the Lyman break galaxies may contribute equally to the comoving SFR density at $z \sim 4$ \citep{Daddi09}. In addition to their unique energetics, they also represent an important stage in massive galaxy formation. Multiple lines of evidence suggest that SMGs are likely the progenitors of massive elliptical galaxies \citep[e.g.,][]{Lilly99,Swinbank06,Aravena10,Lapi11,Hickox12}, which apparently have formed bulk of their stars rapidly at an early epoch \citep[e.g.,][]{Renzini06}. 

Our understanding of this important high-redshift galaxy population are limited by the sensitivity and spatial resolution of current facilities. Gravitational lensing offers an elegant solution by effectively lifting both limiting factors. Also thanks to the negative $K$-correction, it is relatively straightforward to identify strongly lensed SMGs in large area submillimeter surveys. \mk{\citet{Blain96} and \citet{Negrello07} predict that extragalactic sources with 500~\um\ flux density $S_{500} > 100$ mJy are mostly strongly lensed or blended SMGs, nearby late-type galaxies, and radio active galactic nuclei (AGNs). As demonstrated by \citet{Negrello10}, objects in the last two categories can be easily removed using data at other wavelengths, leading to an extremely high success rate in identifying strongly lensed SMGs with this technique (see also \citealt{Vieira10}).} This simple flux selection has produced a few well-studied strongly lensed SMGs (Lockman01 $z = 3.0$: \citealt{Conley11,Riechers11,Scott11}; ID141 $z = 4.2$: \citealt{Cox11,Bussmann11}; HLS~J091828.6$+$514223 $z = 5.2$: \citealt{Combes12}; and \obj\ $z = 3.3$, the subject of this paper), all of which were discovered by the \herschel\footnote{\herschel\ is an ESA space observatory with science instruments provided by European-led Principal Investigator consortia and with important participation from NASA.} Space Observatory \citep{Pilbratt10}. More complex selection processes have been proposed \citep[e.g.,][]{Gonzalez-Nuevo12}, which would allow selecting hundreds of fainter lensed galaxies with \herschel. 

The brightest of the lensed SMGs might also be detected by the \planck\ mission \citep{Collaboration11i}. With $\sim$4\arcmin\ resolutions, such sources are probably blended with fainter sources even in the highest frequency/resolution channels of \planck\ (545 and 857~GHz, or 550 and 350~\um). The Phase-I 130~deg$^2$ of the \herschel-Astrophysical Terahertz Large Area Survey \citep[H-ATLAS;][]{Eales10a} covers 28 \planck\ sources in the \planck\ Early Release Compact Source Catalog \citep[ERCSC;][]{Collaboration11vii}. \mk{\citet{Herranz12} find that sixteen of them are high Galactic latitude cirrus, ten are low-redshift galaxies, and one is resolved into two similarly bright nearby spirals (NGC~3719 and 3720). Only one \planck\ source is dominated by high-redshift galaxies: PLCKERC857 G270.59+58.52 
($S_{550} = 1.4\pm0.6$~Jy, $S_{350} = 2.1\pm0.8$~Jy\footnote{Flux densities are taken from the ERCSC GAUFLUX column. The source is only detected at 545 and 857~GHz by \planck.}). With 18\arcsec, 25\arcsec, and 36\arcsec\ angular resolutions at 250, 350 and 500~\um, respectively, \herschel\ detect 16 objects within a 4.23\arcmin\ radius of the \planck\ position (\planck\ has a full-width-half-maximum [FWHM] resolution of 4.23\arcmin\ at 857\,GHz). There are 15 faint ($S_{\rm 350} \sim 40$~mJy) sources surrounding an unusually bright source (\objfull, hereafter \obj; $S_{350} = 378\pm28$~mJy). Taking into account the differences in the beam size and the filter transmission, the \herschel\ sources account for only $\sim$28\% and $\sim$24\% of the \planck\ flux densities at 545 and 857~GHz, respectively, suggesting that the \planck\ measurements are boosted because of either positive noise spikes \citep[i.e., ``Eddington Bias'';][]{Eddington13} or blending with an over-density of sources that are below the confusion limit of \herschel\ \citep{Negrello05}. The reader is referred to \citet{Herranz12} for a detailed \planck-\herschel\ comparison.}

\obj\ peaks at 350~\um\ in flux density (``350~\um\ peaker''), implying a high photometric redshift given typical dust temperatures. Subsequent detections of multiple carbon-monoxide (CO) lines from this unusually bright object determined a redshift of $z_{\rm CO} = 3.2592\pm0.0010$ \citep[Zspectrometer, CARMA, Z-Spec;][Van der Werf et al. in prep; Riechers et al. in prep]{Harris11}. The high 500~\um\ flux density ($S_{500} = 298\pm24$~mJy), in combination with the confirmed high redshift, makes \obj\ an excellent strong lens candidate. \mk{It is also the only strongly lensed SMG candidate associated with a \planck\ detection in the entire 130~deg$^2$ H-ATLAS Phase-I region. Although the \planck\ detection is partly due to spurious factors (i.e., Eddington bias and/or blending), the confirmation of the lensed nature of the dominating source demonstrates that \planck\ can efficiently identify the brightest lensed SMGs once Galactic cirrus and low-redshift galaxies are removed.}

In this paper we present a detailed multi-wavelength analysis of this \planck-associated SMG. We describe our high-resolution Keck adaptive optics imaging, Submillimeter Array (SMA) and Jansky Very Large Array (JVLA) interferometric observations, and the panchromatic photometry in \S~\ref{sec:obs}. We then perform a joint strong lens modeling at rest-frame 0.5~\um, 200~\um, and CO(1$\to$0) in \S~\ref{sec:model}. In \S~\ref{sec:sed} we derive the intrinsic physical properties of the SMG from its spectral energy distribution (SED). We conclude by discussing the physical properties of \obj\ in the context of unlensed $z > 2$ SMGs (\S~\ref{sec:summary}). Throughout we adopt a $\Lambda$CDM cosmology with $\Omega_{\rm m}=0.3$, $\Omega_\Lambda=0.7$ and $H_0$ = 70 km~s$^{-1}$~Mpc$^{-1}$.

\section{Observations} \label{sec:obs}

\begin{figure*}
\plotone{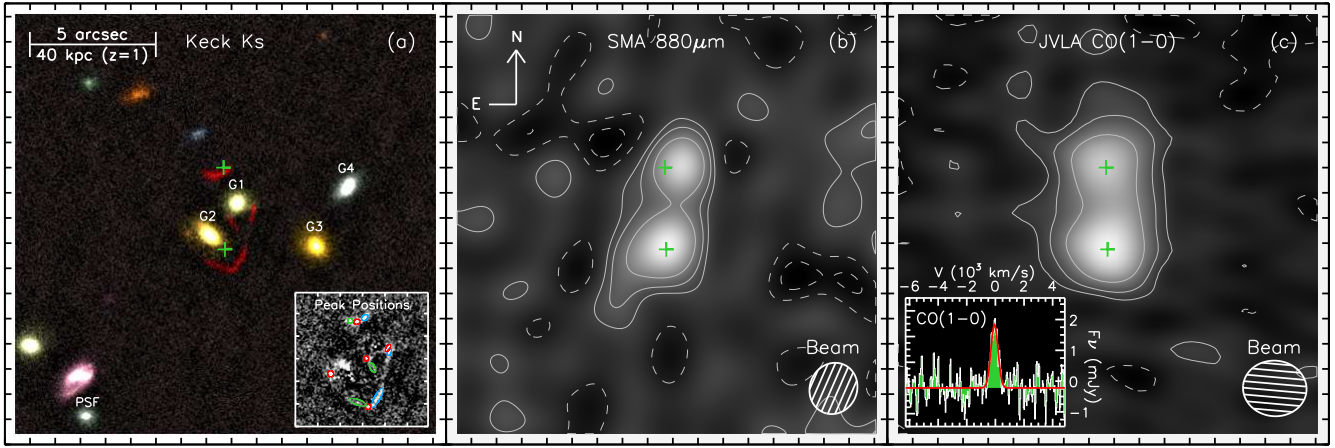}
\caption{High-resolution images of \obj. All images are aligned and the tickmarks are spaced at intervals of 1\arcsec. Green crosses mark the two components seen in the JVLA image. $a$: Keck \K-band image painted with a pseudo-colormap from Keck \K\ (Red), $J$ (Green), and ACAM optical (Blue) images. Lensing galaxies and the PSF star are labelled. The scale bar indicates 5\arcsec\ or 40~kpc at the lens redshift. The inset shows the lens-subtracted \K-band image overlaid with the peak positions for lens modeling (\S~\ref{sec:peak}). For clarity, the positional errors, as indicated by the ellipses, are enlarged by a factor of four. The colors distinguish images from the three clumps in the source plane. 
$b$: SMA 880~\um\ compact array image. Contours are drawn at $-2, -1, +1, +2,$ and $+4\sigma$, where $\sigma$ is the r.m.s.\ noise (3~mJy~beam$^{-1}$).
$c$: JVLA CO(1$\to$0) image. Contours are drawn at $-1, +2, +4,$ and $+8\sigma$, where $\sigma$ is the r.m.s.\ noise (27~\uJy~beam$^{-1}$). The inset shows the CO spectrum from the same data cube, along with a Gaussian fit (red). In $b$ and $c$, the ellipse to the lower right shows the beam.
\label{fig:obs}}
\end{figure*}

\subsection{Keck Adaptive Optics Imaging}

We obtained a 3,440-s $K_{\rm S}$-band (hereafter \K) image on 2011 April 13 (UT) and a 2,100-s $J$-band image on 2011 June 30 (UT) with the Keck\,II laser guide-star adaptive-optics system \citep[LGSAO;][]{Wizinowich06}. An $R = 15.8$ magnitude star 48\arcsec\ SW of \obj\ served as the tip-tilt reference star. The estimated Strehl ratios at the source position are $\sim$23\% and 5\% in \K\ and $J$-band, respectively. We used the NIRC2 camera at 0\farcs04 pixel$^{-1}$ scale for both filters (40\arcsec\ field), and dithered with 2$-$3\arcsec\ steps. The \mk{atmospheric} seeing at 0.5~\um\ was $\sim$0.4\arcsec\ and 0.5\arcsec\ during the \K\ and $J$-band imaging, respectively\footnote{http://kiloaoloa.soest.hawaii.edu/current/seeing/}. 

We used our IDL (Interactive Data Language) programs to reduce the images. After dark subtraction and flat-fielding, sky background and object masks are updated iteratively. For each frame, after subtracting a scaled median sky, the residual background is removed with B-spline models. In the last iteration, we discard the three frames of the poorest image quality and correct the NIRC2 geometric distortion using the solution of P. B. Cameron\footnote{http://www2.keck.hawaii.edu/inst/nirc2/forReDoc/post\_observing/dewarp/} before combining the aligned frames. The resolution of the final \K\ and $J$-band images are 0\farcs16 and 0\farcs27 in FWHM, respectively. We measure the FWHMs from the most compact source in the field located 10\arcsec\ SE of \obj\ (labeled ``PSF'' in Fig.~\ref{fig:obs}$a$); we also use this object as the PSF in the lens modeling (\S~\ref{sec:amoeba}). The images are flux calibrated against UKIRT Infrared Sky Survey \citep[UKIDSS;][]{Lawrence07} and reach depths of \K\ = 25.6 and $J = 25.0$ AB for a 5$\sigma$ detection with 0\farcs1 and 0\farcs2 radius apertures\footnote{Different aperture sizes were chosen here because of the different resolutions.}, respectively. 

\subsection{William Herschel Telescope Imaging} 

Limited by the small field of NIRC2, a deep wide-field image is required for astrometry calibration. Optical imaging was obtained with the high-throughput auxiliary-port camera (ACAM) mounted at a folded-Cassegrain focus of the 4.2-m William Herschel Telescope \citep{Benn08} on 2011 April 26 (UT). We obtained four images of 200~s on a $\sim$2\arcmin\ field centered on \obj, without any filter. The seeing was $\sim$0\farcs9. The images were reduced and combined following standard techniques in {\sc IRAF}\footnote{http://iraf.noao.edu/}. No accurate photometric calibration is possible because we did not use any broad-band filter. But by comparing sources extracted from the ACAM image and the SDSS $i$-band catalog in the same field, we find that our image reaches an equivalent $i$-band 5-$\sigma$ depth of 24.6 AB, or 2.3 magnitudes deeper than the SDSS.

We solve the astrometry of the ACAM image using the on-sky positions of Sloan Digital Sky Survey \citep[SDSS;][]{Aihara11} DR8 sources inside the field. We use the astrometry routines in Marc Buie's IDL library\footnote{http://www.boulder.swri.edu/$\sim$buie/idl/} to correct for offsets, rotation, and distortions, with four terms (constant, $X$, $Y$, and $R = \sqrt{X^2+Y^2}$). Sources that appear blended in the SDSS catalog are excluded. With 35 SDSS sources, we measure 1$\sigma$ dispersions of \dra\ = 0\farcs13 and \ddec\ = 0\farcs14 between the astrometry calibrated ACAM image and the SDSS. Finally, we use the same routines to match the NIRC2 images to the ACAM image with 13 well-detected sources inside the 40\arcsec\ NIRC2 field of view. The corrected NIRC2 images show 1$\sigma$ dispersions of \dra\ = 0\farcs04 and \ddec\ = 0\farcs05.

\subsection{SMA Submillimeter Imaging}

\mk{We obtained SMA interferometric imaging of \obj\ at 880~\um\ (339.58~GHz) in the compact array configuration with an on-source integration time ($t_{\rm int}$) of 1~hr and at 890~\um\ (336.9~GHz) in the subcompact array configuration with $t_{\rm int}$ = 2~hr. The compact and subcompact observations took place on 2011 May 2 and 2012 January 14, respectively. During both nights, atmospheric opacity was low ($\tau_{\rm 225~GHz} \sim 0.1$) and phase stability was good. Both observations used an intermediate frequency coverage of 4--8~GHz and provide a total of 8~GHz bandwidth (considering both sidebands). The quasars 1229$+$020 and 1058$+$015 were used for time-variable gain (amplitude and phase) calibration. The blazar 3C~279 served as the primary bandpass calibrator. For the compact data, we used Titan as the absolute flux calibrator. For the subcompact data, we intended to use Callisto as the flux calibrator, but Jupiter might have fallen into one of the side lobes of the SMA primary beam while we observed Callisto. So we decided to use 3C~279 in lieu of Callisto as the flux calibrator. It is possible to use 3C~279 because we have reliable measurements of its flux both before and after the observation of \obj.}

\mk{We used the {\sc invert} and {\sc clean} tasks in the Multichannel Image Reconstruction, Image Analysis, and Display (MIRIAD) software \citep{Sault95} to invert the {\it uv} visibilities and deconvolve the dirty map, respectively. We used natural weighting to obtain the best sensitivity. For the compact data, the {\sc clean}ed image has a synthesized beam with a FWHM resolution of $2\farcs07\times1\farcs87$ at a position angle (PA) of $-23.6$~degrees east of north; for the subcompact data, the beam is $5\farcs57\times3\farcs68$ at PA = 65.7~deg. The primary beam of the SMA is $\sim$37\arcsec. The r.m.s.\ noise levels are 3.0~mJy~beam$^{-1}$ and 3.6~mJy~beam$^{-1}$ for the compact image and the subcompact image, respectively.}

\mk{\obj\ is resolved into two components by the SMA (Fig.~\ref{fig:obs}). Taking into account the 10\% flux calibration uncertainty, the total flux is $70\pm10$~mJy and $93\pm12$~mJy for the compact image and the subcompact image, respectively. The latter agrees well with the LABOCA bolometer array flux measurement at 870~\um\ (\S~\ref{sec:photometry}). The compact array data did not fully capture the source flux because of the sparser array configuration, i.e., $\sim25\%$ of the total flux is distributed on spatial scales larger than those accessible to the SMA in its compact array configuration. So we use the total flux from subcompact data for SED modeling (\S~\ref{sec:sedsmg}). We chose to use the compact image for lens modeling (\S~\ref{sec:sma}) because of its higher spatial resolution. We find that using the subcompact image or the subcompact$+$compact combined image does not change the lens modeling result, but they give larger errors for the derived parameters.}

\subsection{JVLA CO(1$\to$0) Imaging}

We exploited the recent upgrade to the National Radio Astronomy Observatory\footnote{NRAO is operated by Associated Universities Inc., under a cooperative agreement with the National Science Foundation.}'s Very Large Array \citep{Perley11}, which includes the provision of Ka-band receivers (26.5--40\,GHz), to observe the redshifted CO(1$\to$0) emission from \obj\ at 27.06532\,GHz ($\nu_{\rm rest} = 115.27120$\,GHz; \citealt{Morton94}).

Observations were carried out dynamically during excellent weather conditions on 2012 January 6 and 8. During this Open Shared Risk Observing period the available bandwidth from the new Wideband Interferometric Digial ARchitecture (WIDAR) correlator consisted of two independently tunable output pairs of eight sub-bands each, with $64\times 2$-MHz full-polarisation channels per sub-band, giving a total bandwidth of 2,048\,MHz. At the redshift of \obj, however, the CO(1$\to$0) line could be reached by only the BD output pair, giving $\sim 11,350$\,\kms\ coverage and $\sim 22$\,\kms\ resolution. We offset our tuning by 64\.MHz to avoid noisier edge channels. The 8 sub-bands of output pair AC were tuned to 32.5\,GHz.

The bright compact calibration source, J1150$-$0023 were observed every few minutes to determine accurate complex gain solutions and bandpass corrections. 3C\,286 ($S$ = 2.1666~Jy at 27.06~GHz) was also observed to set the absolute flux scale, and the pointing accuracy was checked locally every hour. In total, around 2\,hr of data were obtained for \obj, with $\sim$1\,hr of calibration.

The data were reduced using {$\cal AIPS\/$} (31DEC12) following the procedures described by \citet{Ivison11}, though with a number of important changes: data were loaded using {\sc bdf2aips} and {\sc fring} was used to optimize the delays, based on 1\,min of data for 3C\,286. The base bands were knitted together using the {\sc noifs} task, yielding $uv$ datasets with 512 $\times$ 2-MHz channels, which we then added together using the task {\sc dbcon}. Finally, the channels were imaged over a $512\times 512\times 0.3''$ field, with natural weighting ({\sc robust = 5}), to form a $512^3$ cube centered on \obj. Integrating over those 55 channels found to contain line emission (so a {\sc fwzi} of $\sim 1,200$\,\kms) yielded an r.m.s.\ noise level of 27\,$\mu$Jy\,beam$^{-1}$. 

The {\sc clean}ed and velocity-integrated CO map is shown in Fig.~\ref{fig:obs}$c$. The beam is 2\farcs5$\times$2\farcs2 at PA = $85^{\circ}$. Similar to the SMA, the map resolves two components separated by $\sim$5\arcsec. The CO lines extracted from the two components show the same redshift and line profile, further confirming that they are lensed images of a single source. The best-fit Gaussian to the area-integrated spectrum gives a line width of $\Delta V_{\rm FWHM}=585\pm55$\,\kms\ and a line flux of $S_{\rm CO}\Delta V = 1.52\pm0.20$\,Jy\,\kms. In comparison, the CO(1$\to$0) measurements reported by \citet{Harris11} using Zpectrometer on the Green Bank Telescope are: $\Delta V_{\rm FWHM}=680\pm80$\,\kms\ and $S_{\rm CO}\Delta V = 1.18\pm0.26$\,Jy\,\kms (corrected for the 20\% difference in the absolute flux density of 3C~286). \mk{The reason for the discrepancy is unclear, but the two line flux measurements agree within the 1$\sigma$ errors. So hereafter, we use the weighted mean of the two measurements, $S_{\rm CO}\Delta V = 1.40\pm0.22$\,Jy\,\kms, to derive the molecular gas mass.} 

\subsection{Panchromatic Photometry} \label{sec:photometry}

\begin{deluxetable}{llccc} 
\tablewidth{0pt}
\tablecaption{Photometry \label{tab:photo}}
\tablehead{ 
\colhead{Instrument} & \colhead{Band} & \colhead{$\lambda$} & \colhead{$F_{\nu}$(G1+G2)} & \colhead{$F_{\nu}$(SMG)} \\
\colhead{} & \colhead{} & \colhead{(\um)} & \colhead{(\uJy)} & \colhead{(mJy)}
}
\startdata
SDSS     & $u$    &     0.36& $0.1\pm0.1$ & \nodata           \\
         & $g$    &     0.47& $0.4\pm0.4$ & \nodata           \\
         & $r$    &     0.62& $3.3\pm1.2$ & \nodata           \\
         & $i$    &     0.75& $8.6\pm2.0$ & \nodata           \\
         & $z$    &     0.89& $ 23\pm8$   & \nodata           \\
UKIDSS   & $Y$    &     1.03& $ 35\pm6$   & \nodata           \\
         & $J$    &     1.25& $ 67\pm8$   & $0.0017\pm0.0003$ \\
         & $H$    &     1.63& $ 68\pm10$  & \nodata           \\
         & $K$    &     2.20& $139\pm10$  & $0.0123\pm0.0009$ \\
WISE     & $w1$   &     3.35& $205\pm20$  & $0.037\pm0.020$   \\
         & $w2$   &     4.60& $242\pm53$  & $\lesssim0.117$          \\
         & $w3$   &    11.56& $735\pm35$  & $\lesssim0.702$          \\
         & $w4$   &    22.09& $ <3460$    & \nodata           \\
PACS     & green  &      100& \nodata     & $ 25\pm6$         \\
         & red    &      160& \nodata     & $138\pm21$        \\
SPIRE    & blue   &      250& \nodata     & $323\pm24$        \\
         & green  &      350& \nodata     & $378\pm28$        \\
         & red    &      500& \nodata     & $298\pm24$        \\
LABOCA   &\nodata &      870& \nodata     & $103\pm19$        \\
SMA      &\nodata &      890& \nodata     & $ 93\pm12$        \\
MAMBO    &\nodata &    1200 & \nodata     & $ 38\pm6$         \\
CARMA    &\nodata &    2792 & \nodata     & $1.4\pm0.5$       \\
         &\nodata &    3722 & \nodata     & $     <2.0$       \\
VLA      &\nodata &  214000 & \nodata     & $1.2\pm0.4$       \\
\enddata
\end{deluxetable}

Photometry of \obj\ were obtained from the SDSS ($u, g, r, i,$ and $z$), the UKIDSS ($Y, J, H,$ and $K$), the Wide-Field Infrared Survey Explorer \citep[\wise, 3.6 and 4.6~\um;][]{Wright10}, the \herschel/PACS \citep[100 and 160~\um; Program ID: OT1\_RIVISON\_1;][]{Ibar10}, the \herschel/SPIRE \citep[250, 350, and 500~\um;][]{Pascale11, Rigby11}, the Large APEX BOlometer CAmera \citep[LABOCA, 870~\um;][]{Siringo09}, the SMA (880~\um), the Max-Planck Millimetre Bolometer \citep[MAMBO, 1.2 mm;][]{Kreysa99}, the Combined Array for Research in Millimeter-wave Astronomy \citep[CARMA, 2792 and 3722~\um;][]{Bock06}, and the VLA FIRST survey \citep[21 cm;][]{Becker95}. 

We obtained imaging at 870~\um\ with the LABOCA bolometer array at the Atacama Pathfinder EXperiment (APEX) telescope in November 2011 (Clements et al. in prep). LABOCA observed a 11.4\arcmin\ diameter field with a resolution of FWHM = 18\farcs6. The observations have a total integration time of $\sim$30~hr reaching a 1$\sigma$ sensitivity of $\sim$2~mJy.
 
We obtained 1.2~mm imaging with MAMBO at the IRAM 30-m telescope (FWHM $\sim$ 10\farcs7) in January and February 2011 (Dannerbauer et al. in prep.). Observing time in the on-off mode is 24~minutes, achieving a 1$\sigma$ sensitivity of $\sim$1~mJy.

We obtained continuum observations at 81.2 and 108.2~GHz (3722 and 2792~\um; covering rest-frame CO[3$\to$2] and CO[4$\to$3] lines) on 2011 Mar 18 and Sep 1 as part of our CO follow-up campaign of bright, lensed H-ATLAS SMGs with CARMA in D array (Riechers et al. in prep). Observations were carried out for 0.9 and 1.4~hr on source, respectively, using the 3~mm receivers and a bandwidth of 3.7~GHz per sideband. \obj\ is unresolved in these observations, with angular resolutions of $6\farcs8\times5\farcs0$ and $6\farcs0\times3\farcs8$ at 81.2 and 108.2~GHz, respectively (restored with natural baseline weighting).

Table~\ref{tab:photo} lists the photometry. We have included in the errors the absolute flux calibration uncertainties (3\% for \wise, 3$-$5\% for PACS, 7\% for SPIRE, 10\% for SMA, and 15\% for LABOCA, MAMBO, and CARMA).

\section{Lens Modeling} \label{sec:model}

Because the LGSAO image has the highest spatial resolution, we use it to find the best-fit lens model. We initially use the peak positions of the multiply-imaged source to constrain the lensing potentials (\S~\ref{sec:peak}), then we exploit the $K$-band light distribution in the image plane to quantify the morphologies of the source as well as refining the lensing potentials (\S~\ref{sec:amoeba}). Finally, we use the best-fit lensing potentials and the SMA and JVLA images to constrain the sizes and locations of the dust and molecular gas in the source plane (\S~\ref{sec:sma} \& \ref{sec:evla}).

\subsection{\K-band Peak Positions} \label{sec:peak}

We use {\sc lenstool} \citep{Kneib96,Jullo07} to find the best-fit parameters and their errors from the peak positions. {\sc lenstool} implements a Bayesian Markov chain Monte-Carlo sampler to derive the posterior distribution of each parameter and an estimate of the evidence for the model. 

The lensing system is mainly made of two red filaments that are $\sim$3\farcs5 apart (Fig.~\ref{fig:obs}$a$). The outward curved shape of the northern arc can be explained if the source is intrinsically curved. Hence we split each of the two arcs into three parts and build a simple lens model by putting two deflectors centered on G1 and G2. We find that the predicted counter-images can explain the additional features close to G1 and G2. Guided by the predicted counter-images, we define three systems of lensed images (Fig.~\ref{fig:obs}$a$ $inset$). The 11 peak positions in 3 separate systems provide a total of 16 constraints ($11\times2-3\times2$), allowing us to include shear from nearby galaxies G3 and G4.

For the lensing galaxies, we find photometric redshifts of $z_{\rm G1+G2} = 1.06\pm0.16$ and $z_{\rm G4} = 0.80\pm0.28$ with the public photo-$z$ code {\sc EAZY} \citep{Brammer08}. We obtain the nine-band photometry from the SDSS ($u, g, r, i,$ and $z$) and the UKIDSS ($Y, J, H,$ and $K$) surveys. At these wavelengths, the flux from the lensed galaxy is negligible (\S~\ref{sec:sed}). \mk{The redshift of G1$+$G2 is measured from the total fluxes of G1 and G2, because they are blended in the seeing limited data.} G3 is undetected in SDSS but shows similar color as G1 and G2. Hence in the lens modeling we assume all four galaxies are at $z = 1.06$. \mk{Note that although redshift errors of the lensing galaxies would lead to errors in the estimated lens masses, they would not change our conclusions on the lensed galaxy because the magnification factors would remain the same.} 

For the lens model, we assume that the dark-matter plus baryonic mass profiles of the foreground lens galaxies G1 to G4 can be described as singular isothermal ellipsoids \citep[SIEs;][]{Kormann94}. The SIE profile is parametrized by the velocity dispersion ($\sigma$), the position ($x$, $y$), the axis ratio ($q = b/a$), and the PA ($\theta$, E of N). We fix the positions to the centers of the galaxies. For G3 and G4, we further fix their $q$ and $\theta$ to those from the light distribution, because they are not well constrained by the peak positions and there are significant correlations between the PA and ellipticity of the light and of the mass distribution \citep[e.g.][]{Sluse11}. Therefore, we have a total of eight free parameters. We find a best-fit with $\chi^2 = 7.9$ for dof = 8 (degrees of freedom) and an average positional error of 0\farcs04 ($\sim$1 pixel). The parameters and their errors are summarized in Table~\ref{tab:lens}. We also list the mass enclosed by the critical curve for each SIE, 
\begin{equation}
M_E = \frac{4\pi^2}{G} \frac{D_L D_{LS}}{D_S} \frac{\sigma^4}{c^2}, \label{eq:ME}
\end{equation}
where $D_L$, $D_S$, and $D_{LS}$ are the angular diameter distances to the lens, to the source, and between the lens and the source, respectively. The radius of the area enclosed by the critical curve can be approximated by the circularized Einstein radius: 
\begin{equation}
b = 4.5 (\frac{\sigma}{200~{\rm km}~\rm{s}^{-1}})^2 \sqrt{\frac{2q}{1+q^2}}~{\rm kpc}. \label{eq:b}
\end{equation} 
In the errors of masses and velocity dispersions, we have included the 1-$\sigma$ uncertainty of the photometric redshift. 

The nominal model described above is the \mk{most favorable} description of the lensing system because of the following: 
\begin{enumerate}

\item Adding $q$'s and $\theta$'s of G3 and G4 as free parameters does not substantially improve the fit: the Bayesian evidence\footnote{The improvement of a model is substantial if $1 < \Delta\ln(E) < 2.5$, strong if $2.5 < \Delta\ln(E) < 5$, and decisive if $\Delta\ln(E) > 5$ \citep{Jeffreys61}.} increases by only $\Delta\ln(E)$ = 0.5 and the reduced $\chi^2$ actually increases from 1.0 to 1.6 as a result of the decreased degree of freedom. 

\item Excluding the potentials of G3 and/or G4 does degrade the fit significantly. The Bayesian evidence decreases by $\Delta\ln(E)$ = 2.5 and 30, and the reduced $\chi^2$ increases from 1.0 to 1.8 and 6.5, when we exclude G4 and both G3 and G4, respectively. 

\item Including a group-scale potential with a PIEMD profile \citep[Pseudo-Isothermal Elliptic Mass Distribution;][]{Kassiola93} does not improve the fit. For the PIEMD profile, we adopt a cut-off radius of 500~kpc but allow the position, ellipticity, PA, core radius, and velocity dispersion to vary. To limit the number of free parameters, we fix the $q$'s and PA's of the SIEs to those measured from the lensing galaxies but allow their velocity dispersions to vary. So we have a total of 10 free parameters. The reduced $\chi^2$ of the best-fit is 1.9, much higher than that of the nominal model. The Bayesian evidence also decreases by $\Delta\ln(E)$ = 6.9 when compared with the nominal model.

\end{enumerate}

\begin{deluxetable}{llccr} 
\tablewidth{0pt}
\tablecaption{Derived Properties from Lens Modeling
\label{tab:lens}}
\tablehead{ 
\colhead{Object} & \colhead{Quantity} & \multicolumn{2}{c}{Value} & \colhead{Unit} \\
\colhead{} & \colhead{} & \colhead{Peak Position$^a$} & \colhead{Surface Brightness$^b$} & 
}
\startdata
G1     &$         q$ & $ 0.46\pm 0.10$ &$ 0.52\pm 0.04$ &\nodata\\
\nodata&$    \theta$ & $  -59\pm   10$ &$  -61\pm    2$ &deg\\
\nodata&$    \sigma$ & $  209\pm   24$ &$  212\pm   24$ &\kms\\
\nodata&$       M^c$ & $ 1.6_{ -0.6}^{+0.9}\times10^{11}$ &$ 1.7_{ -0.7}^{+0.9}\times10^{11}$ &\msun\\
G2&$         q$ & $ 0.63\pm 0.06$ &$ 0.69\pm 0.03$ &\nodata\\
\nodata&$    \theta$ & $   59\pm    7$ &$   67\pm    2$ &deg\\
\nodata&$    \sigma$ & $  240\pm   28$ &$  240\pm   27$ &\kms\\
\nodata&$       M^c$ & $ 2.7_{ -1.1}^{+1.5}\times10^{11}$ &$ 2.7_{ -1.1}^{+1.5}\times10^{11}$ &\msun\\
G3&$    \sigma$ & $  243\pm   51$ &$  242\pm   28$ &\kms\\
\nodata&$       M^c$ & $ 2.9_{ -1.8}^{+3.3}\times10^{11}$ &$ 2.8_{ -1.1}^{+1.5}\times10^{11}$ &\msun\\
G4&$    \sigma$ & $  168\pm   40$ &$  165\pm   20$ &\kms\\
\nodata&$       M^c$ & $ 6.6_{ -4.4}^{+8.9}\times10^{10}$ &$ 6.1_{ -2.4}^{+3.6}\times10^{10}$ &\msun\\
 & & & & \\
SMG     & $\mu(K)$          & \nodata & \muKeck & \nodata \\
\nodata & $\mu({880\mu m})$ & \nodata & \muSMA  & \nodata \\
\nodata & $\mu({\rm CO})$   & \nodata & \muJVLA & \nodata \\
\enddata
\tablecomments{
$a$ - Best-fit parameters from the \K-band peak positions (\S~\ref{sec:peak}).
$b$ - Best-fit parameters from the \K-band surface brightness distribution (\S~\ref{sec:amoeba}).
$c$ - Total mass enclosed by the critical curve of each SIE potential (Eqs.~\ref{eq:ME} \& \ref{eq:b}). 
}
\end{deluxetable}

\subsection{\K-band Source} \label{sec:amoeba}

Although the peak positions can constrain the deflectors through ray tracing, they cannot provide an accurate estimate of the magnification factor because the source-plane light distribution is not taken into account. Because we want to estimate the intrinsic properties of the lensed galaxy, we are interested in the luminosity weighted magnification factor, which depends on the source morphology because the magnification factor is different at each source plane position. In this section we model the morphology of the lensed galaxy and refining the lensing potentials simultaneously with the \K-band image. The PSF is derived from the most compact source in the field, which is 10\arcsec\ SE of \obj\ (Fig.~\ref{fig:obs}$a$). 

Following \S~\ref{sec:peak}, we assume that the source consists of three clumps, each described as a \sersic\ profile. Again we use SIE profiles for the lensing potentials. So we have a total of 29 parameters: seven parameters for each \sersic\ profile, and eight parameters for the SIE potentials. Our fitting procedure is as follows. For an initial set of parameters describing the source and the lenses from \S~\ref{sec:peak}, we use {\sc lenstool} to generate lensed images of the source, which is then convolved with the PSF and compared with the observed image. We limit the comparison in a $3\farcs3\times4\farcs5$ ($83\times113$~pixels) rectangular region that encloses the lensing features. This process is iterated with AMOEBA\_SA to find the parameters that minimize the residual between the observation and the model. AMOEBA\_SA is based on the IDL multidimensional minimization routine AMOEBA \citep{Press92} with simulated annealing added by E.~Rosolowsky. We allow a maximum of 1000 iterations in each call of AMOEBA\_SA. For the simulated annealing, we adopt an initial ``temperature'' of 100 and decrease it by 20\% in each subsequent call to AMOEBA\_SA. A good fit with a reduced $\chi^2$ around unity is normally found after a few calls to AMOEBA\_SA. For each iteration, we compute the total luminosity-weighted magnification factor ($\mu_{K}$) by summing the pixel values in the image and the source planes with apertures matched by inverting the image plane aperture to the source plane. The 1-$\sigma$ confidence interval of $\mu_{K}$ is found with $\chi^2(\mu)-\chi^2_{\rm min} \leq 1$. Note that we compute the $\chi^2$ values on the residual image binned by $4\time4$-pixel boxes (FWHM = 0\farcs16 = 4 pixel), so that the noise becomes uncorrelated between pixels; or equivalently, one could divide the $\chi^2$ values from the original residual images by a factor of 16. We find the luminosity-weighted magnification to be $\mu_{K} = $ \muKeck. The best-fit parameters for the deflectors are listed in Table~\ref{tab:lens}. The results are very similar to those from fitting the peak positions, although the errors are smaller because the entire image provides more information than the peak positions alone.

\begin{figure*}
\plotone{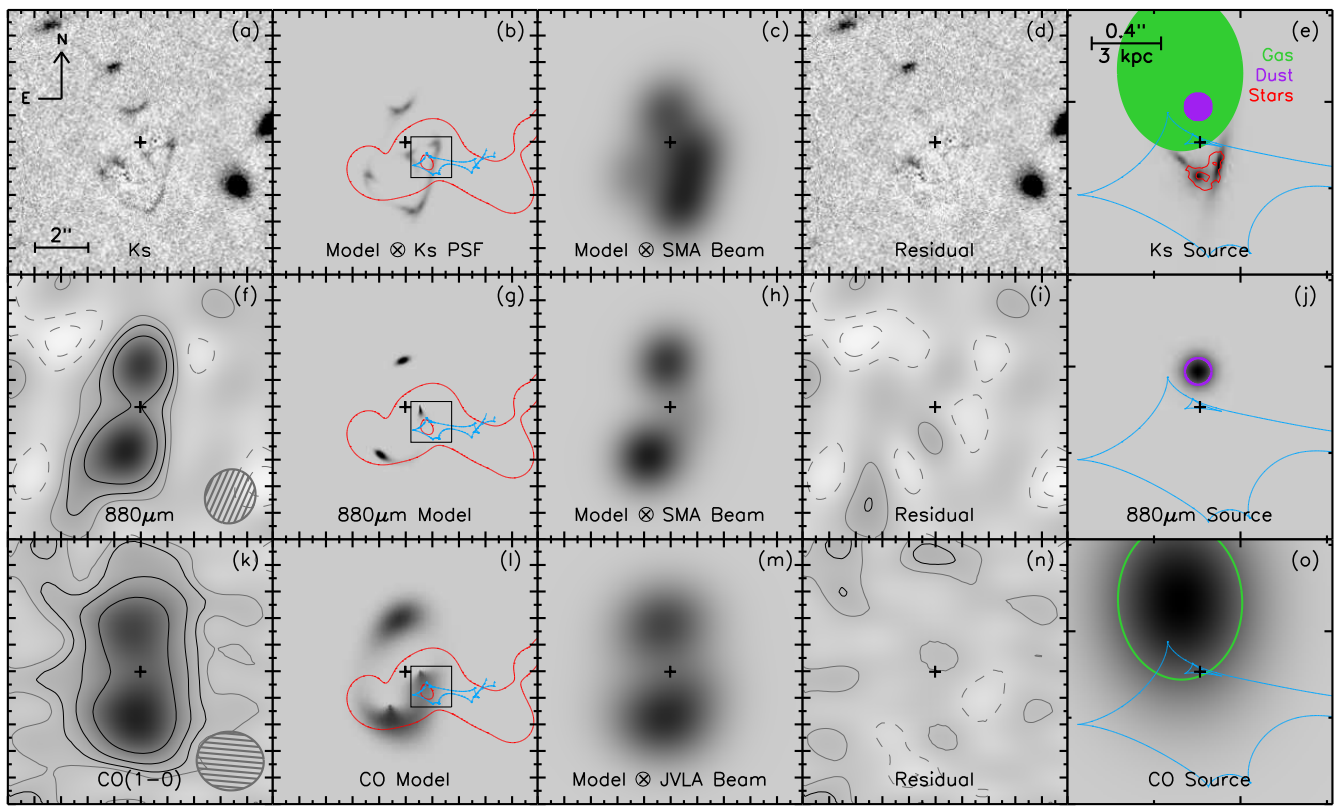}
\caption{Lens modeling results. Major tickmarks are spaced at intervals of 1\arcsec. To ease comparisons, a cross is drawn at the center of each panel.
$a$: Keck \K\ image after subtracting G1 and G2. 
$b$: Best-fit \K\ model convolved with the \K-band PSF. Critical curves are in red and caustics are in blue. The box delineates the region covered by the source images (i.e., $e, j$, and $o$).
$c$: \K\ model convolved with the SMA beam. It is clearly different from the SMA and JVLA images ($f$ \& $k$), indicating differential magnification.
$d$: \K\ residual.
$e$: Modeled intrinsic source morphology (i.e., without PSF; grey scale) vs. a direct inversion of the observed image (red contours). For comparison, the 880~\um\ ($purple$) and CO(1$\to$0) ($green$) sources are shown as color-filled ellipses.  
$f$: SMA 880~\um\ compact array image. The grey ellipse shows the beam. Here and in $i$, contours are drawn at $-2, -1, +1, +2,$ and $+4\sigma$, where $\sigma$ is the r.m.s.\ noise (3~mJy~beam$^{-1}$).
$g$: 880~\um\ model.
$h$: Model convolved with the SMA beam.
$i$: 880~\um\ residual.
$j$: 880~\um\ source. The purple circle shows the FWHM of the source.
$k$: JVLA CO(1$\to$0) image. Here and in $n$, contours are drawn at $-1, +1, +2, +4,$ and $+8\sigma$, where $\sigma$ is the r.m.s.\ noise (27~\uJy~beam$^{-1}$).
$l$: CO model.
$m$: Model convolved with the JVLA beam.
$n$: CO residual.
$o$: CO source. The green ellipse shows the FWHMs of the source.
\label{fig:lens}}
\end{figure*}

Figure~\ref{fig:lens} shows the best-fit model. The lensed galaxy has a curved morphology, causing the northern arc bending in the opposite direction of the deflectors. In the source plane, the three clumps extend over only 0\farcs21 or 1.6~kpc, and their effective radii are $0\farcs21\pm0\farcs04$ ($1.5\pm0.3$ kpc), $0\farcs085\pm0\farcs013$ ($0.6\pm0.1$ kpc), and $0\farcs11\pm0\farcs05$ ($0.8\pm0.4$ kpc) from W to E.

The nature of the feature NNE of G2 in the residual image is unclear, but it is unlikely to be at the same redshift as the lensed galaxy: tracing its position to the source plane and imaging it back predicts an \emph{unobserved} equally bright counter-image 0\farcs8 S of the southern arc. This feature could therefore be part of the galaxy G2.

\begin{figure*}
\plottwo{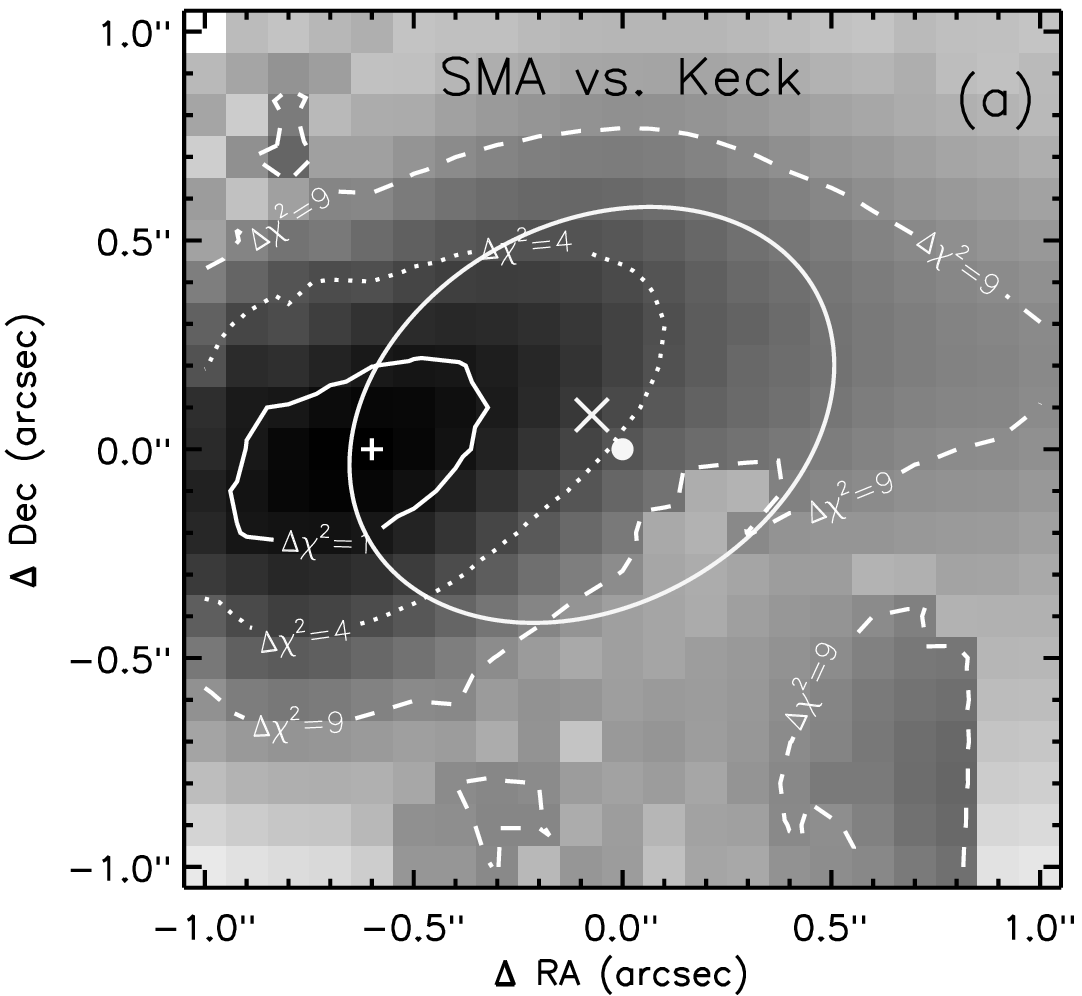}{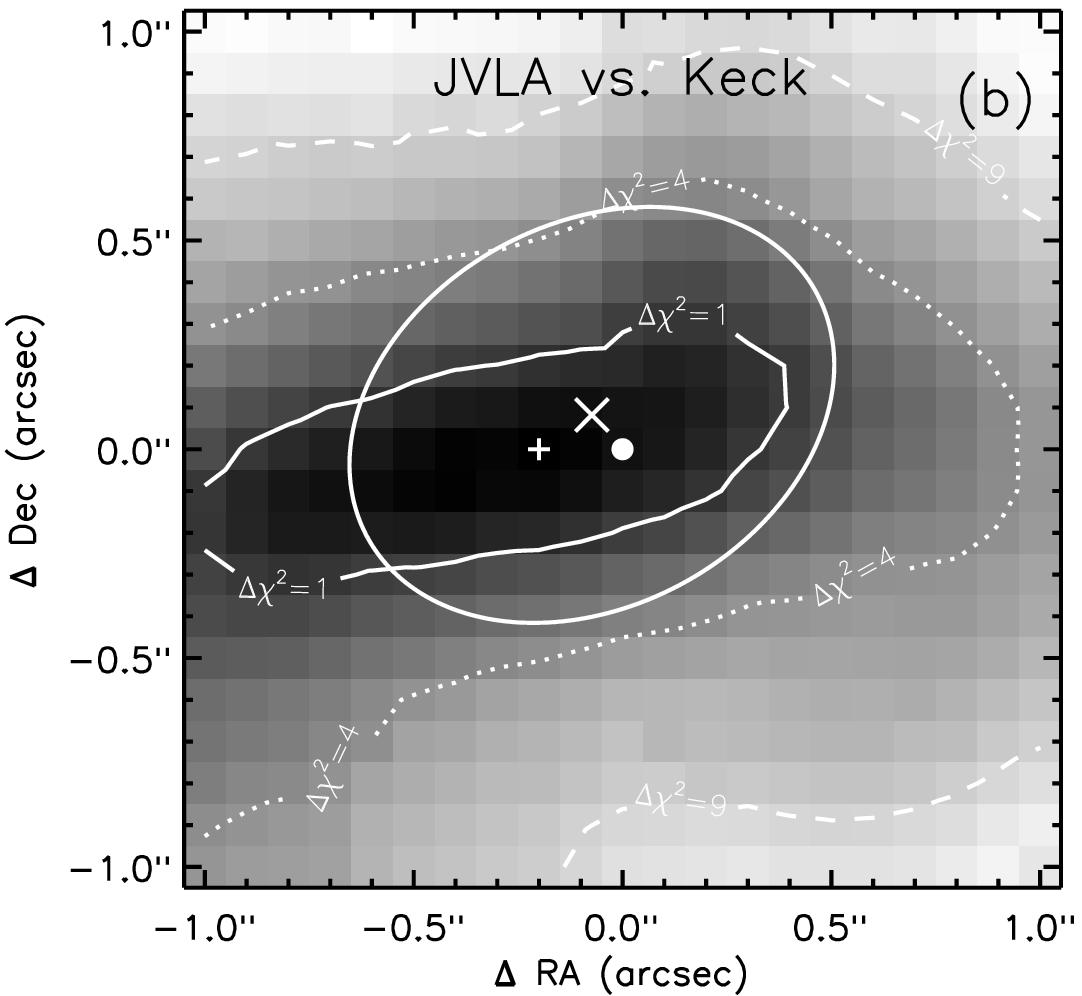}
\caption{Constraining the astrometric offset between SMA, JVLA and Keck. $a$: The background image is the $\chi^2$ map of the best-fit models as a function of SMA$-$Keck offset. The image is displayed in logarithmic scale. Positive offsets indicate shifting the SMA image W or N relative to the Keck image. Iso-$\chi^2$ contours are overlaid for 1, 2, and 3$\sigma$ intervals. The white plus sign indicates the offset that yields the minimum $\chi^2$ value. The ellipse shows the 1$\sigma$ astrometry uncertainty determined from FIRST--SDSS cross-correlation. The cross indicates the systematic offset ($-$0\farcs07, $+$0\farcs08) between FIRST and SDSS within $1^{\circ}$ radius of \obj. The overlapping area between the 1$\sigma$ contour of the $\chi^2$ map and the ellipse gives the best estimate of the astrometric offset and its uncertainty. Zero offset is indicated by the white circle. $b$: Same as $a$ but for JVLA relative to Keck.
\label{fig:xychi2}}
\end{figure*}

\subsection{880~\um\ Source} \label{sec:sma}

Precise astrometry calibration is crucial for a joint analysis of images from different wavelengths. Because the only \K\ source detected by the SMA is \obj, we have to estimate the astrometry offset between the two images in a statistical way. Because the Keck image is tied to the SDSS astrometry and the SMA image is tied to the radio reference frame, we cross-correlate the VLA FIRST catalog \citep{Becker95} and the SDSS catalog within $1^{\circ}$ of \obj\ and compute the optical-radio separation. Ninety-four radio sources have optical counterparts within 3\arcsec. We then fit an elliptical Gaussian to the two-dimensional distribution in $\Delta {\rm RA} = -(\alpha_{\rm FIRST} - \alpha_{\rm SDSS})$ and $\Delta {\rm Dec} = \delta_{\rm FIRST} - \delta_{\rm SDSS}$. The systematic offset from the peak position of the Gaussian is consistent with zero ($\Delta {\rm RA} = -0\farcs07$, $\Delta {\rm Dec} = +0\farcs08$). The best-fit Gaussian has $\sigma$'s of 0\farcs40 and 0\farcs30, and a PA of $119^{\circ}$ for the major axis. Therefore, the 1$\sigma$ ellipse of the astrometry offset has major/minor semi-axes of 0\farcs61/0\farcs45. Our result is consistent with that of \citet{Ivezic02}, who found a $\sim$0\farcs1 systematic offset and a 1$\sigma$ error circle of 0\farcs47 in radius between FIRST and SDSS astrometry. 

We can constrain the astrometry offset further through lens modeling. As demonstrated by \citet{Kochanek92} and \citet{Wucknitz04}, interferometric data are most naturally modeled with the $uv$-plane visibilities, because it avoids beam deconvolution and naturally handles correlated noise. Here, however, we opt to model the {\sc clean}ed map directly, because (1) the images are essentially unresolved in the SMA map, and (2) we already have a good lens model from the \K-band image (\S~\ref{sec:amoeba}). Because of the limited spatial resolution of the SMA 880~\um\ image, the two centroid positions do not offer enough information to constrain the lens model. Hence, for the deflectors, we fix all the parameters to the best-fit values from \S~\ref{sec:amoeba}; for the source, we assume a circular Gaussian profile with variable position and size. We shift the SMA image relative to the \K\ model on a 2\arcsec$\times$2\arcsec\ grid with 0\farcs1 steps. At each offset position, we find the best-fit model using the same fitting procedure as in \S~\ref{sec:amoeba}. The modeling is performed on a $51\times68$-pixel ($5\farcs1\times6\farcs8$) region enclosing the SMA sources. Figure~\ref{fig:xychi2}$a$ shows a map of the minimum $\chi^2$ values at each offset position. The global best-fit, with reduced $\chi^2$ of unity, is reached when we shift the SMA image $0\farcs6$ E of the \K\ image. The middle panels of Fig.~\ref{fig:lens} show this global best-fit model. 

The noise of the SMA map is Gaussian but is highly correlated. We compute the r.m.s.\ noise of the SMA map after binning it by boxes of $n^2$ pixels. We find that the noise starts to decrease as $1/n$ for $n \gtrsim 20$~pixels (FWHM $\simeq$ 2\arcsec\ = 20 pixel), indicating that the noise becomes uncorrelated on 20-pixel scales. Therefore, we divide the $\chi^2$ values from the the residual images by a factor of 400, which is equivalent to computing $\chi^2$ from residual images binned by $20\time20$-pixel boxes.

In combination with the 1$\sigma$ error ellipse from FIRST--SDSS cross-correlation, we determine that the astrometry offset between 880~\um\ and \K\ images is $\Delta {\rm RA} = -0\farcs5\pm0\farcs1$ and $\Delta {\rm Dec} = 0\farcs0\pm0\farcs2$; i.e., the overlapping region between the ellipse and the 1$\sigma$ contour of the $\chi^2$ map. Collecting all of the solutions in this permitted offset region satisfying $\chi^2(\mu)-\chi^2_{\rm min} \leq 1$, we estimate a luminosity-weighted 880~\um\ magnification of $\mu_{\rm 880} =$ \muSMA, and an 880~\um\ source size of FWHM = $0\farcs15^{+0.14}_{-0.06}$ = $1.2^{+1.0}_{-0.5}$~kpc. Because we have fixed the deflectors with the best-fit parameters from \K-band, the errors here do not include the uncertainties of the deflectors. Higher resolution far-IR images are required to constrain the deflectors and the source simultaneously. 

Dust emitting regions are often spatially offset from the UV/optical emitting regions in SMGs \citep{Tacconi08, Bothwell10, Carilli10, Ivison10b, Riechers10}. This is clearly the case for \obj, which shows distinctly different morphologies at \K-band and 880~\um, even after convolving the \K-band image with the SMA beam (compare Fig.~\ref{fig:lens}$c$ \& $f$). From the lens model, we estimate a source-plane separation between the 880~\um\ source and the central \K\ clump of $0\farcs41\pm0\farcs07$ or $3.1\pm0.5$~kpc (Fig.~\ref{fig:lens}$e$).

If we assume zero astrometry offset between SMA and Keck, then we obtain a model that poorly fits the observation ($\Delta\chi^2 \sim 4$; Fig.~\ref{fig:xychi2}$a$). The lens model gives a slightly larger magnification ($\mu_{\rm 880} = 8.4\pm1.6$) and doubles the source size (FWHM = $2.5^{+1.9}_{-0.3}$~kpc). However, the source-plane separation between the 880~\um\ source and the central \K\ clump remains the same ($3.2\pm0.2$~kpc).

\subsection{CO(1$\to$0) Source} \label{sec:evla}

We use the same technique to model the JVLA CO(1$\to$0) map as in \S~\ref{sec:sma}. The lensed images are better resolved in the JVLA image than in the SMA image, so we use an elliptical Gaussian instead of a circular Gaussian for the source profile. The model has a total of six free parameters ($x, y$, FWHM, $q$, PA, and flux density). Again, we can constrain the astrometric offset between JVLA and Keck through lens modeling. Figure~\ref{fig:xychi2}$b$ shows the minimum $\chi^2$ values at each offset position relative to the \K\ image. To deal with the correlated noise, we scale the $\chi^2$ values from the residual map by the product of the FWHMs of the major and minor axes of the beam. The global best-fit, with reduced $\chi^2$ of unity, is reached when we shift the JVLA image 0\farcs2 E of the \K\ image. The bottom panels of Fig.~\ref{fig:lens} show the best-fit model. We estimate a CO magnification of $\mu_{\rm CO} =$ \muJVLA. Similar to the SMA image, we find a source-plane separation of $4.7\pm1.6$~kpc between the cold molecular gas and the stellar emission (i.e., the central \K\ clump). The CO(1$\to$0) is emitted from a more extended region than the dust, but the two spatially overlap (Fig.~\ref{fig:lens}$e$). The CO source has FWHM = $0\farcs9\pm0\farcs3$ = $6.8\pm2.3$~kpc along the major axis, with an axis ratio of $0.8^{+0.2}_{-0.6}$. 

The molecular gas disk is massive. The velocity-area-integrated CO brightness temperature of $L'_{\rm CO} = [6.4\pm1.0]\times10^{11}$\,K~km~s$^{-1}$~pc$^2$ indicates a molecular gas reservoir of $M_{\rm gas} = [7.4\pm2.1]\times10^{10}$\,\msun\ after lensing correction, assuming a conversion factor of $\alpha_{\rm CO} = M_{\rm gas}/L'_{\rm CO} = 0.8$ \msun~(K~km~s$^{-1}$~pc$^2$)$^{-1}$, which is commonly assumed for starburst environments \citep{Solomon05}. Note, however, that $\alpha_{\rm CO}$ is uncertain by at least a factor of a few \mk{and it may depend on the metallicity, the gas temperature, and the velocity dispersion of the galaxy \citep{Narayanan12}}. 

The CO magnification factor determined from the lens model (\muJVLA) is in excellent agreement with that estimated from the CO luminosity$-$FWHM correlation. The observed $L'_{\rm CO}$ and line width indicates a magnification factor of $7\pm2$, based on its deviation from the correlation established by unlensed SMGs \citep[][Bothwell et al. in prep]{Harris11}. This agreement demonstrates that strongly lensed SMGs may be effectively selected with CO spectroscopy in the future.

\section{Spectral Energy Distributions} \label{sec:sed}

\begin{figure*}
\plottwo{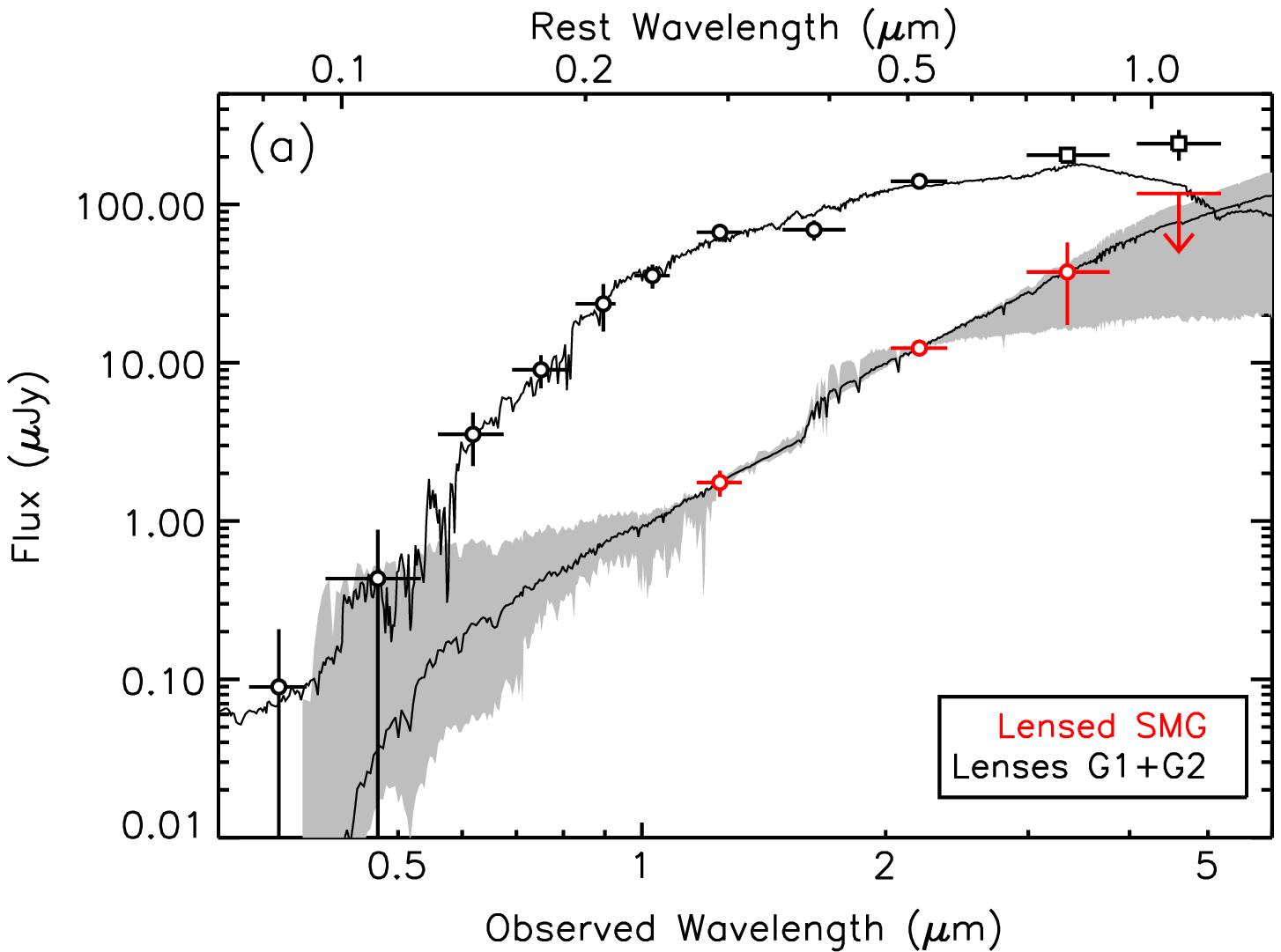}{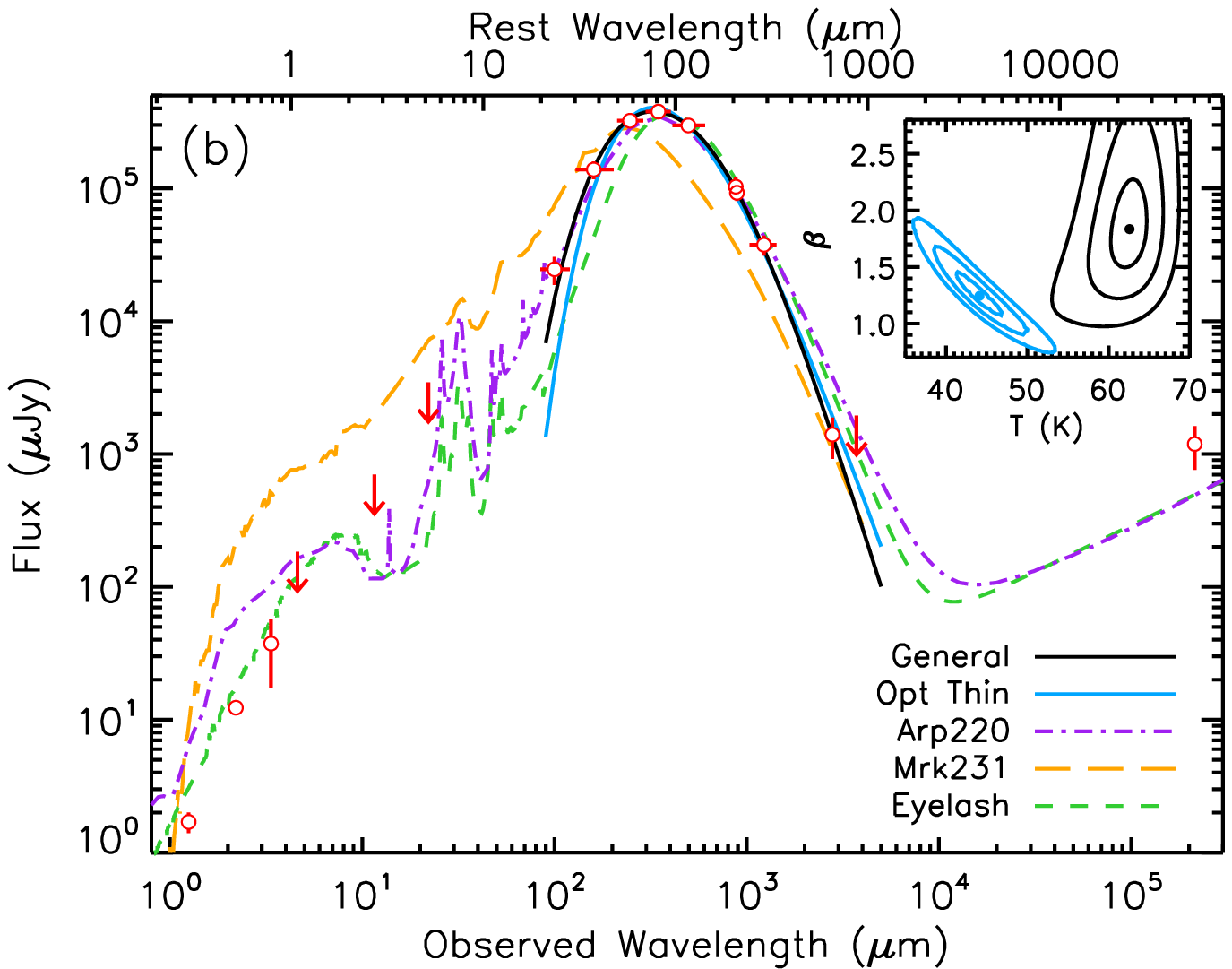}
\caption{Modeling the SEDs. Top axes indicate wavelengths at the rest frame of the SMG ($z = 3.26$). $a$: Black data points are for the foreground lenses G1$+$G2, and the red data points are for the lensed SMG. The top black curve shows the best-fit BC03 stellar population synthesis model of G1$+$G2 at $z = 1.06$, using the nine data points below 3~\um. The bottom black curve shows the best-fit BC03 model for the SMG, along with the 1$\sigma$ range of acceptable models. $b$: The full SED of the SMG. The short dashed ($green$), dash-doted ($purple$), and long dashed ($orange$) curves are the best-fit SED templates of the ``Cosmic Eyelash'' \citep{Ivison10}, Arp~220, and Mrk~231, respectively. The Eyelash provides the best description of the overall SED among the three. The solid black and blue curves are the best-fit models with a single-temperature modified blackbody using the general and optically thin formula, respectively. The inset shows the 1, 2, and 3$\sigma$ contours in the $T-\beta$ plane for the general ($black$) and optically thin ($blue$) models.
\label{fig:sed}}
\end{figure*}

\begin{deluxetable}{llcr} 
\tablewidth{0pt}
\tablecaption{Derived Physical Properties
\label{tab:sedfit}}
\tablehead{ 
\colhead{Object} & \colhead{Quantity} & \colhead{Value} & \colhead{Unit}
}
\startdata
G1$+$G2 & $M_{\rm stellar}$ & $3.5^{+1.8}_{-1.1}\times10^{11}$ & \msun \\
\nodata & Age & $3\pm2$ & Gyr \\
\nodata & $\tau$ & $0.3\pm0.2$ & Gyr \\
\nodata & $E(B-V)$ & $0.04^{+0.11}_{-0.04}$ & mag \\
\nodata & SFR$_{\rm opt}$ & $0.1^{+0.4}_{-0.1}$ & \msunyr \\
& & & \\
SMG$^a$ & $M_{\rm stellar}$    & $[3.5\pm2.4]\times10^{10}$ & \msun \\
\nodata & SFR$_{\rm opt}$      & $1000\pm1000$ & \msunyr \\
& & & \\
SMG$^b$ & $T_{\rm dust}$       & $63\pm2$ & K \\
\nodata & $\beta$              & $1.8\pm0.4$ & \nodata \\
\nodata & $M_{\rm dust}$       & $[7.0\pm2.0]\times10^8$ & \msun \\
\nodata & $\lambda_0$          & $250\pm40$  & \um \\
\nodata & $\sigma$             & $0.8\pm0.2$ & kpc$^2$ \\
\nodata & $L_{\rm IR}$         & $[1.7\pm0.3]\times10^{13}$ &  \lsun \\
\nodata & SFR$_{\rm IR}$       & $1900\pm400$ & \msunyr \\
& & & \\
SMG$^c$ & $T_{\rm dust}$       & $44\pm3$ & K \\
\nodata & $\beta$              & $1.3\pm0.2$ & \nodata \\
\nodata & $M_{\rm dust}$       & $[1.0\pm0.3]\times10^9$ & \msun \\
& & & \\
SMG$^d$ & $L'_{\rm CO}$        & $[9.3\pm2.6]\times10^{10}$ & K~km~s$^{-1}$~pc$^2$ \\
\nodata & $L_{\rm IR}/L'_{\rm CO}$   & $172\pm58$ & \lsun/K~km~s$^{-1}$~pc$^2$ \\
\nodata & FWHM$_{\rm CO}$      & $585\pm55$ & \kms \\
\nodata & $M_{\rm gas}$      & $[7.4\pm2.1]\times10^{10}$ & \msun \\
\nodata & $M_{\rm dyn}$        & $[3.2\pm1.3]\times10^{11}$ & \msun \\
\nodata & $M_{\rm gas}/M_{\rm baryon}$ & $68\pm17$\% & \nodata \\
\nodata & $M_{\rm gas}/M_{\rm dyn}$ & $23\pm11$\% & \nodata \\
\enddata
\tablecomments{
Magnification-dependent parameters have been demagnified and their errors include the magnification uncertainties. 
$a$ - Stellar population synthesis modeling of the near-IR SED.
$b$ - General ``optically thick'' modified blackbody fit to the far-IR-to-mm SED. 
$c$ - Optically thin modified blackbody fit. 
$d$ - Parameters derived from CO(1$\to$0) observations.
}
\end{deluxetable}

Useful physical parameters are encoded in the SEDs. The optical-to-NIR SED of \obj\ is dominated by the foreground galaxies G1 and G2, from which we can derive the photometric redshift and the stellar population of the lensing galaxies. The far-IR and submillimeter regime is dominated by the lensed SMG, as evident in the SMA image, so the data can tell us the dust and star formation properties of the SMG.  

\subsection{Lensing Galaxies} \label{sec:sedlens}

Adopting the photometric redshift of 1.06, we model the nine-band photometry ($u$, $g$, $r$, $i$, $z$, $Y$, $J$, $H$, and $K$) of G1+G2 with the stellar population synthesis models of \citet[][BC03]{Bruzual03}. We assume a \citet{Chabrier03} initial mass function (IMF), \citet{Calzetti94} extinction law, and exponentially declining star formation history, with a range of e-folding times ($\tau$ = 0.1 to 30~Gyr) and ages (0.01 to 12.5~Gyr). For each template, we fit for the stellar mass ($M_{\rm stellar}$) and extinction ($E(B-V)$). The best-fit model gives $\chi^2 = 6.4$ for dof = 7 (Fig.~\ref{fig:sed}$a$). The derived properties of G1$+$G2 are listed in Table~\ref{tab:sedfit}. The intrinsic extinction is small ($E(B-V) = 0.04^{+0.11}_{-0.04}$) and there is very little current star formation (SFR = $0.1^{+0.4}_{-0.1}$ \msunyr). The dust-absorbed UV/optical luminosity ($[3^{+14}_{-3}]\times10^{10}$ \lsun) is less than 0.15\% of the total integrated IR luminosity before lensing correction (\lir\ = $1.2\times10^{14}$ \lsun). Therefore, G1 and G2 do not contribute significantly to the far-IR fluxes, in agreement with their absence in the SMA image. The stellar mass from SED modeling is $\sim$80\% of the total mass within the critical curves from lens modeling (Table~\ref{tab:sedfit}), implying that the galaxies are dominated by stellar mass within $\sim7$~kpc.

\subsection{Lensed SMG} \label{sec:sedsmg}

In the $J$ and $K$-bands, we obtain the photometry of the SMG with an aperture contoured around the multiply-imaged features after subtracting the foreground lenses. We measure $K = 21.2\pm0.1$ and $(J-K) = 2.1\pm0.2$ in AB magnitudes, consistent with the red $J-K$ colors of unlensed SMGs \citep{Frayer04,Dannerbauer04}. Careful modeling is required to extract SMG photometry from the \wise\ data, because the SMG is blended with the foreground galaxies G1 through G4 (FWHM = 6$-$12\arcsec). We model the \wise\ 3.4~\um\ source with four elliptical Gaussians of the same shape. With {\sc galfit} \citep{Peng10}, we fix their positions to those determined from the Keck image, but we allow the Gaussian shape to vary. Then we measure the flux density of G1$+$G2 ($\sim$205~\uJy) decomposed from that of G3 and G4 ($\sim$117~\uJy). Finally, the flux density of the SMG ($\sim$37~\uJy) is estimated from the excess of G1$+$G2 relative to their best-fit stellar population synthesis model (\S~\ref{sec:sedlens}). Unfortunately, we can not separate G3$+$G4 from G1$+$G2 in the longer wavelength channels of \wise\ because of the inferior image quality, so we treat the excesses over the best-fit model of G1$+$G2 as upper limits for the SMG. The far-IR-to-millimeter SED is dominated by the lensed SMG, therefore no foreground subtraction is necessary. Table~\ref{tab:photo} summarizes the photometry for \obj.

We opt to model the rest-frame optical and far-IR emission separately, instead of fitting them together in a self-consistent way with {\sc magphys} \citep[][]{Cunha08}, because our lens model shows that they are emitted from physically distinctive regions, i.e., the dust that attenuates the optical emission has little to do with the starburst-heated dust that emits in the far-IR. 

It is difficult to constrain the stellar population with only three photometric detections (dof = 1) in rest-frame optical. However, we can limit the parameter space by excluding unphysical models, such as those that require negative extinction corrections and those whose ages exceed the cosmic age at $z = 3.2592$ (1.9~Gyr). We use the same BC03 templates as in the previous section. The shaded region in Fig.~\ref{fig:sed}$a$ shows all of the permitted models with $\chi^2 < {\rm dof}+1$. These models give a range of extinctions, stellar masses, and SFRs: $E(B-V) < 0.94$, $M_{\rm stellar} = [3.5\pm2.4]\times10^{10}$ \msun, and SFR $< 2000$ \msunyr\ (Table~\ref{tab:sedfit}). The dust-absorbed UV/optical luminosity range from 0\% to 140\% of the observed \lir; but 90\% of the models have dust-absorbed luminosity less than 50\% of the observed \lir. 

We fit the far-IR SED with a single-temperature modified blackbody, 
	\begin{equation}
	S(\nu_{\rm obs}) = \sigma~(1-e^{-\tau})~B(\nu_{\rm rest},T)~(1+z)~\mu/d_L^2
	\end{equation}
where $\sigma$ is the total absorption cross section of dust particles at the optically thick limit (i.e., the size of the dust-obscured region), $B(\nu,T)$ the \planck\ function, $\tau = (\nu_{\rm rest}/\nu_0)^{\beta} = (\lambda_0/\lambda_{\rm rest})^{\beta}$ the optical depth, $\mu$ the lensing magnification factor, and $d_L$ the luminosity distance. In the optically thin limit ($\lambda \gg \lambda_0$), dust mass can be derived based on the knowledge of the opacity $\kappa_d$ (absorption cross section per unit mass): 
	\begin{equation}
	M_{\rm dust} = \frac{S(\nu_{\rm obs})~d_L^2}{\kappa_d(\nu_{\rm rest})~B(\nu_{\rm rest},T)~(1+z)~\mu}
	\end{equation} 
It is generally assumed that the opacity follows a power law, $\kappa_d(\nu) \propto \nu^{\beta}$, and has a normalization of $\kappa_d = 0.07\pm0.02$~m$^2$~kg$^{-1}$ at 850~\um\ \citep{Dunne00,James02}. Both the general ``optically thick'' ($S_{\nu} \propto (1-e^{-\tau}) B_{\nu}(T)$) and the optically thin ($S_{\nu} \propto \nu^\beta B_{\nu}(T)$) models provide good fits to the observed SED (Fig.~\ref{fig:sed}$b$). For the general model, we use all of the nine detections between 100~\um\ and 3~mm. \mk{The best-fit general model gives $\chi^2 = 1.6$ for dof = 5, suggesting that the photometric errors have been overestimated. For the optically thin model, we exclude the PACS 100~\um\ point, which is clearly on the Wien tail where small grains tend to dominate the emission. The best-fit optically thin model gives $\chi^2 = 4.0$ for dof = 5. The derived parameters are listed in Table~\ref{tab:sedfit}.}

The optically thick model yields dust properties similar to those of the local Ultra-Luminous Infrared Galaxy (ULIRG) Arp~220 \citep{Rangwala11}, with the optical depth exceeds unity below rest-frame $\sim$250~\um. The intrinsic $8-1000$~\um\ luminosity of $L_{\rm 8-1000} = [1.7\pm0.3]\times10^{13}$ \lsun\ classifies \obj\ as a hyper-luminous infrared galaxy (Hy-LIRG). The IR luminosity implies an SFR of $1900\pm400$ \msun~yr$^{-1}$ for a \citet{Chabrier03} IMF \citep{Kennicutt98}. Using the values of $L_{\rm 8-1000}$ and $T = 63\pm2$~K in the Stefan-Boltzmann law we obtain a spherical source radius of $780\pm100$~pc, which is three times larger than that of Arp~220 (230~pc) because of the ten times greater luminosity. The source radius is comparable to that we derive from the optically thick model ($r = 500\pm60$~pc) and is consistent with the size we measure from modeling the SMA image (FWHM = $1.2^{+1.0}_{-0.5}$~kpc; \S~\ref{sec:sma}). Therefore, the optically thick model is preferred. 

The radio luminosity from the observed 1.4~GHz flux density is $L_{\rm 1.4GHz} = 4 \pi d_L^2 S_{\rm 1.4GHz} (1+z)^{\alpha-1} = [7.3\pm3.4]\times10^{25}$~W~Hz$^{-1}$ for a radio spectral index of $\alpha = 0.7$. \mk{Assuming the radio emission is magnified by the same factor as the submillimeter emission, the IR-to-radio luminosity ratio of \obj, $q_L = {\rm log}(L_{\rm IR}/({\rm 4.52 THz}~L_{\rm 1.4GHz})) = 2.1\pm0.2$, is consistent with the radio--far-IR correlation of high-redshift starburst galaxies: e.g., \citet{Kovacs06} measured $q_L = 2.14\pm0.12$ for 15 SMGs, while \citet{Ivison10c} measured $q_L = 2.40\pm0.24$ for 65 \herschel\ 250~\um\ selected galaxies. This suggests that the AGN contribution is insignificant in \obj.} 

We also do not see significant AGN contribution in the mid-IR. In Fig.~\ref{fig:sed}$b$, we fit the SEDs of the local ULIRGs Mrk~231, Arp~220, and the $z = 2.3$ SMG ``Cosmic Eyelash'' \citep{Ivison10} to the far-IR SED. The \wise\ upper limits lie well below the AGN-dominated ULIRG Mrk~231 but are more consistent with Arp~220 and the Eyelash. Therefore, we conclude that \obj\ is predominantly a starburst system. 

\section{Discussion and Conclusions} \label{sec:summary}

We have presented high-resolution \K-band, 880~\um, and CO(1$\to$0) observations and the near-IR-to-centimeter SED of a \herschel-selected strongly lensed SMG at $z = 3.2592$ (\obj). The SMG shows distinctly different morphologies in the three images, suggesting differential magnification due to stratified morphologies. A joint strong lens modeling shows that the SMG is lensed by four galaxies at $z \sim 1$ and the luminosity-weighted magnification factors are \muKeck\ in \K, \muSMA\ at 880~\um, and \muJVLA\ at CO(1$\to$0). In the source plane, the SMG consists of several stellar clumps extended over $\sim$1.6~kpc with $[3.5\pm2.4]\times10^{10}$~\msun\ of stars, a compact ($\sim$1~kpc) starburst enshrouded by $[7.0\pm2.0]\times10^8$ of dust at $\sim$60\,K, and an extended ($\sim$6~kpc) cold molecular gas reservoir with $[7.4\pm2.1]\times10^{10}$~\msun\ of gas. The starburst and its gas reservoir are located $\sim$4~kpc from the stars. \mk{Similar separations between optical and submillimeter/radio emission have been observed in unlensed high-redshift dusty starbursts \citep[e.g.,][]{Tacconi08, Casey09, Bothwell10, Carilli10, Riechers10}. However, these previous results could also be attributed to astrometry offsets across different facilities; \citet{Casey09} might be the only exception, whose tied the astrometry of the images to larger radio and optical fields. In \obj,} the physical separations amongst stars, dust, and gas are less ambiguous because of the clear wavelength-dependent morphologies in the image plane: fortuitously, the stars and dust/gas straddle across the caustic, so the less obscured \K-band region is quadruply imaged while the heavily obscured starburst and its gas reservoir are doubly imaged. Because of the $\sim$4~kpc separation between the stars and the gas-rich starburst and their similar masses, it is tempting to suggest that the SMG is in the process of a major merger, which presumably is driving the starburst activity in $z > 2$ SMGs \citep[e.g.,][]{Tacconi08,Engel10}. However, spatial separation of this scale can also be explained by differential dust obscuration in a single galaxy, as has been proposed for other high-redshift SMGs where spatial offsets have been observed between rest-frame UV and submillimeter \citep[e.g., GN20 and AzTEC~3;][]{Carilli10,Riechers10}. 

\mk{Differential magnification may affect the observed far-IR SED as well as the CO ladder, as hotter dust and higher-$J$ CO lines may have more compact morphologies \citep{Ivison11}. However, because the magnification map is smooth in the area, the six times difference in the sizes of the CO(1$\to$0) and the dust emitting region only lead to a $\sim$10\% difference in magnification, which is smaller than the 1$\sigma$ errors of our estimates of the magnification factors. Therefore, differential magnification is unlikely to be significant enough to affect the far-IR SED and future CO ladder measurements in \obj.}  

How does the massive gaseous disk compare with the disks in other SMGs? We estimate an dynamical mass of $[3.2\pm1.3]\times10^{11}$\,\msun\ and a gas fraction of $f_{\rm gas} = M_{\rm gas}/M_{\rm dyn} = 23\pm11$\% for the CO(1$\to$0) disk using the ``isotropic virial estimator'' \citep[e.g.,][]{Tacconi08}:
\begin{equation}
M_{\rm dyn} = 2.8\times10^5 \Delta V_{\rm FWHM}^2 r_{\rm HWHM}~{\rm M}_{\odot},
\end{equation}
where $\Delta V_{\rm FWHM}$ is the CO line FWHM in \kms\ and $r_{\rm HWHM}$ is half of the FWHM size of the disk in kpc. Combined with the FWHM disk radius, we further estimate a gas surface density of $\Sigma = 510\pm370$~\msun~pc$^{-2}$. Both $f_{\rm gas}$ and $\Sigma$ are similar to those of the extended CO(1$\to$0) disks in the two $z \sim 3.4$ SMGs in \citet[][]{Riechers11b}. But both values are significantly smaller than those of the kinematically resolved CO(6$\to$5) disk of the ``Cosmic Eyelash'' \citep[$f_{\rm gas} \sim 70\%$, $\Sigma \sim 3000\pm500$\,\msun\,pc$^{-2}$;][]{Swinbank11}. The discrepancies illustrate the limitations of these widely used but crude estimators {\it and/or} that high excitation CO lines probe more compact and denser regions in a disk. Higher resolution observations are clearly needed to resolve this issue. Without spatially resolved gas kinematics, we refrain from estimating the disk stability parameter of \citet{Toomre64}. 

\obj\ is a gas-rich, initial starburst system similar to unlensed SMGs and local ULIRGs. Its intrinsic IR luminosity well exceeds $10^{13}$ \lsun\ (i.e., Hy-LIRG), implying an enormous rate of star formation ($1900\pm400$ \msun~yr$^{-1}$). Although the molecular gas reservoir is massive and it constitutes $68\pm17$\% of the \emph{visible} baryonic mass ($M_{\rm gas} + M_{\rm dust} + M_{\rm stellar}$) and $23\pm11$\% of the dynamical mass, it will exhaust in just $39\pm14~(\alpha_{\rm CO}/0.8)$~Myr at the current SFR (assuming no gas accretion). The star formation timescale, $\tau_{\rm SF} = M_{\rm stellar}$/SFR = $18\pm13$ Myr, is only $\sim$1\% of the cosmic age at $z = 3.2592$ ($\tau_{\rm cosmic} = 1.9$~Gyr), suggesting that \obj\ is an initial starburst system with maturity $\mu = \tau_{\rm SF}/\tau_{\rm cosmic} \ll 1$ \citep{Scoville07c}. The specific SFR, sSFR = SFR/$M_{\rm stellar}$ = $54\pm38$ Gyr$^{-1}$, is consistent with the average $z > 2$ SMGs, but it is an order of magnitude higher than the median value of the star-forming main sequence of Lyman break galaxies at the same epoch \citep[][]{Daddi09}. 
\mk{We can also estimate the star formation efficiency: $\epsilon = t_{\rm dyn}/(M_{\rm gas}/{\rm SFR})$, where $t_{\rm dyn} = \sqrt{r^3/(2GM)}$ is the dynamical or free-fall timescale. For $r = r_{\rm HWHM}$ = 3~kpc and $M = M_{\rm dyn}$, we obtain $\epsilon = 0.11\pm0.04$, or [11$\pm$4]\% per dynamical timescale, which is comparable to unlensed SMGs but is an order of magnitude higher than normal starforming galaxies \citep[][]{Genzel10}.}

The dust mass of $[7.0\pm2.0]\times10^{8}$~\msun\ is similar to the average dust mass of unlensed SMGs \citep[e.g.,][]{Michalowski10a}, and the gas-to-dust ratio, $M_{\rm gas}/M_{\rm dust} = 110\pm40~(\kappa_{\rm 850\mu m}$/0.07~m$^2$~kg$^{-1})~(\alpha_{\rm CO}$/0.8), is comparable to that of the Milky Way. Assuming that the dust emission is indicative of the size of the starburst and a starburst disk radius of $r_0 = 1$ kpc, the star formation surface density of $\dot{\Sigma}_{\star} \simeq 600\pm120$ \msunyr~kpc$^{-2}$ approaches the Eddington limit of radiation pressure supported starburst disks \citep[$\dot{\Sigma}_{\star} \sim 10^3$ \msunyr~kpc$^{-2}$;][]{Scoville03b,Thompson05}, \mk{similar to local ULIRGs such as Arp~220 and the host galaxy of the $z = 6.4$ quasar SDSS~J114816.64+525150.3 \citep{Walter09b}}.

In conclusion, \obj\ is a \emph{bona fide} SMG with an intrinsic submillimeter flux density of $S_{\rm 880} = 9.2\pm2.2$~mJy. The starburst disk, where most of the molecular gas and dust reside, is spatially separated from the less obscured stellar population by $\sim$4~kpc, suggesting either a major merger or differentiated dust obscuration. The $\sim$1~kpc radius starburst disk is presumably supported in large by radiation pressure on the dust grains. Its physical properties, such as molecular gas mass, stellar mass, gas-to-dust ratio, gas fraction, SFR, star formation efficiency, and radio-to-far-IR luminosity ratio, are all very similar to unlensed $z > 2$ SMGs \citep{Hainline11,Wardlow11,Michalowski10a,Kovacs06}. The lensing boost of the effective angular resolution and sensitivity has allowed us to examine in unprecedented details the properties of a typical starburst galaxy when the Universe is only 1/7 of its current age. \obj\ provides a prelude to a golden age of SMG research, as \herschel\ is unveiling hundreds of strongly lensed SMGs before the mission completes \citep[e.g.,][Wardlow et al. in prep]{Gonzalez-Nuevo12}.

\acknowledgments
We thank the anonymous referee for comments that helped improve the paper.
HF, AC, JLW and SK acknowledge support from NSF CAREER AST-0645427.
IPF is supported by the Spanish grants ESP2007-65812-C02-02 and AYA2010-21697-C05-04.
SGD acknowledges a partial support from the NSF grant AST-0909182.
GDZ and MN acknowledge support from ASI/INAF agreement I/072/09/0 (``\textit{Planck} LFI Activity of Phase E2'') and from MIUR through the PRIN 2009.
Some of the data presented herein were obtained at the W.M. Keck Observatory, which is operated as a scientific partnership among the California Institute of Technology, the University of California and the National Aeronautics and Space Administration. The Observatory was made possible by the generous financial support of the W.M. Keck Foundation. The authors wish to recognize and acknowledge the very significant cultural role and reverence that the summit of Mauna Kea has always had within the indigenous Hawaiian community.  We are most fortunate to have the opportunity to conduct observations from this mountain.
The \herschel-ATLAS is a project with \herschel, which is an ESA space
observatory with science instruments provided by European-led
Principal Investigator consortia and with important participation from
NASA. The H-ATLAS website is http://www.h-atlas.org/. The US
participants acknowledge support from the NASA \herschel Science Center/JPL.
Partly based on observations obtained with \planck\ (http://www.esa.int/Planck), an ESA science mission with instruments and contributions directly funded by ESA Member States, NASA, and Canada.
Support for CARMA construction was derived from the Gordon and Betty Moore Foundation, the Kenneth T. and Eileen L. Norris Foundation, the James S. McDonnell Foundation, the Associates of the California Institute of Technology, the University of Chicago, the states of California, Illinois, and Maryland, and the National Science Foundation. Ongoing CARMA development and operations are supported by the National Science Foundation under a cooperative agreement, and by the CARMA partner universities.

{\it Facilities}: Keck:II (LGSAO/NIRC2), SMA, JVLA, WHT, Sloan, UKIDSS, WISE, Herschel, IRAM 30m, APEX, CARMA

\end{document}